\documentclass[aps,twocolumn,superscriptaddress,showkeys]{revtex4-2}
\usepackage{amsmath}
\usepackage{color,xcolor}
\usepackage{graphicx}
\newcommand{\bastar}{\begin{eqnarray*}}
\newcommand{\eastar}{\end{eqnarray*}}
\newskip\humongous \humongous=0pt plus 1000pt minus 1000pt

\newif\ifdtup

\relax
\newcommand{\W}{{\vec W}}
\newcommand{\n}{\hat n}

\newcommand{\hD}{{\hat D}}
\newcommand{\bD}{{\bar D}}
\newcommand{\cD}{{\cal D}}
\newcommand{\pd}{\partial}

\newcommand{\hB}{{\hat B}}

\newcommand{\bA}{{\bar A}}

\newcommand{\bF}{{\bar F}}
\newcommand{\G}{{\vec G}}
\newcommand{\B}{{\vec B}}

\newcommand{\hG}{{\hat G}}

\newcommand{\mn}{{\mu\nu}}
\newcommand{\om}{\omega}

\newcommand{\lam}{\lambda}

\newcommand{\be}{{\bar e}}

\newcommand{\vsig}{{\vec \sigma}}
\newcommand{\nn}{\nonumber}

\newcommand{\cL}{{\cal L}}

\begin{document}
\title{Theory of Magnon Spintronics: Non-Abelian 
Gauge Theory of Electron Spintronics}

\author{Y. M. Cho}
\email{ymcho0416@gmail.com}
\affiliation{School of Physics and Astronomy,
Seoul National University, Seoul 08826, Korea}
\affiliation{Center for Quantum Spacetime, 
Sogang University, Seoul 04107, Korea}  
\author{Franklin H. Cho}
\email{cho.franklin@qns.science}
\affiliation{Center for Quantum Nano Science,
Ewha Woman's University, Seoul 03766, Korea}

\begin{abstract}
Treating the electron as a charged spinon we 
propose a theory of magnon spintronics, 
a non-Abelian gauge theory of SU(2)xU(1), which could be viewed as an effective theory of electron spintronics. Just like QED the theory has the U(1) electromagnetic interaction, but the new ingredient is the non-Abelian SU(2) gauge interaction of 
the magnon with the spinon. A remarkable feature 
of the theory is the photon-magnon mixing,
the mixing between the electromagnetic U(1) gauge 
boson and the diagonal part of the SU(2) magnon gauge boson. After the mixing we have the massless 
Abelian magnon and a massive photon, and 
the doubly charged massive off-diagonal non-Abelian magnons which induce the spin-flip interaction 
to the spinon. The theory is characterized by three scales. In addition to the correlation length fixed 
by the mass of the Higgs field it has two different penetration lengths, the one fixed by the mass of 
the photon which generates the well known Meissner effect and the other fixed by the mass of 
the off-diagonal magnons which generates the non-Abelian Meissner effect. The non-Abelian structure of the theory naturally accommodates 
new topological objects, the non-Abrikosov 
quantized magnonic vortex and non-Abelian magnonic monopole of the Cho-Maison type, as well as the well known Abrikosov vortex. We discuss the physical implications of the non-Abelian gauge theory of 
the magnon spintronics.
\end{abstract}


\maketitle

{\bf Introduction}---Since Bloch predicted 
the existence of the spin wave as the propagation 
of a disturbance of the magnetic ordering in magnetic materials, the spin wave and the spin 
wave quantum (the magnon) have played 
increasingly important roles in condensed matter physics \cite{bloch}. A wide varieties of linear 
and non-pinear spin wave phenomena have increased 
the interest in the spin wave and magnon, particularly because they are considered as potential data carrier for computing devices 
these days. 

Under this background the magnon spintronics, 
the field of spintronics on the structures, 
devices, and circuits which use the spin current carried by the magnons, has become a very 
important part of condensed matter physics. Analogous to the electric current, the magnon-based spin current can be used to carry, transport, 
and process informations. In particular, the use 
of the magnons allows the implementation of novel wave-based computing technologies free from 
the drawbacks inherent to modern electronics 
such as the dissipation of energy due to the Ohmic losses. Because of this and other advantages 
there have been a huge amount of literatures on 
the phenomenologies and applications of the magnon spintronics \cite{chumak,jmmm,hb}. On the other hand, there have been not so many studies which could explain the underlying physics of the magnon spintronics so far \cite{krug,serga,lenk,stamp}.    

{\it The purpose of this Letter is to discuss 
a possible theory of electron spintronics, 
an $SU(2) \times U(1)$ gauge theory of magnon and photon, which could describe the underlying physics 
of the electron spintronics. Viewing the electron 
as a charged $SU(2)$ spinon doublet, we construct 
a non-Abelian gauge theory of magnon interacting with the charged spinon. The theory has two conserved currents, the electromagnetic current 
and the spin current, which play the essential role in the magnon spintronics. The two conserved currents comes from the two Abelian gauge group 
of the theory, the electromagnetic $U(1)$ and 
the $U(1)$ subgroup of the non-Abelian $SU(2)$.
The theory is made of the massive photon, 
a massless Abelian magnon, two massive doubly charged non-Abelian magnons, and a Higgs scalar. 
The theory has the following characteristic features. The photon-magnon mixing which generates 
the massless Abelian magnon responsible for a long range magnetic ordering in the spintronics
and makes the non-Abelian magnon electrically charged, and the Higgs mechanism which makes 
the photon and the non-Abelian magnon massive without any spontaneous symmetry breaking.}

A salient feature of the theory is that 
the spin-spin interaction is described by 
the exchange of the messenger particles, i.e.,
the magnons. Traditionally the spin-spin 
interaction has always been described as 
an instantaneous action at a distance. As far as 
we understand, there has been no self-consistent field theory of spintronics in which the spin-spin interaction is described by the exchange of 
the messenger particle in terms of the Feynman diagrams. This theory does that.
 
{\bf Non-Abelian gauge theory of electron spintronics}---To construct such a theory we first identify the electron as a charged $SU(2)$ spinon doublet $\phi=(\phi_{\uparrow},\phi_{\downarrow})$ which has the electromagnetic $U(1)$ interaction 
and the non-Abelian $SU(2)$ magnon interaction 
which flips the electron spin, and consider 
the following $SU(2)\times U(1)$ gauge theory described by the Lagrangian  
\begin{gather}
\cL =-|{\cal D}_\mu \phi|^2 -\frac{\lambda}{2}\big(|\phi|^2
-\frac{\mu^2}{\lambda}\big)^2
-\frac14 F_\mn^2-\frac14 \G_\mn^2, \nn \\
\cD_\mu \phi =\big(\pd_\mu-i\frac{g}{2} A_\mu
-i\frac{g'}{2} \vsig \cdot \B_\mu \big) \phi
=(D_\mu -i\frac{g}{2} A_\mu) \phi, \nn\\
D_\mu \phi=(\pd_\mu
-i\frac{g'}{2} \vsig \cdot \B_\mu \big) \phi,
\label{lag0}
\end{gather}
where $A_\mu$ and $\B_\mu$ are the U(1)  electromagnetic gauge potential and the SU(2) 
magnon gauge potential, $F_\mn$ and $\G_\mn$ are 
the corresponding field strengths, $g$ and $g'$ 
are the coupling constants. Notice that 
the electron doublet carries both the electric charge $g/2=e$ and spin charge $g'$. 
The justification of the Lagrangian as 
the Lagrangian for the electron spintronics is 
that it has all necessary ingredients for 
the magnon spintronics, photon, magnon, and electron. Moreover, it has the magnonic as well as 
the electromagnetic interaction acting on 
the electron.

To understand the physical content of the Lagrangian,
we need the Abelian decomposition of the non-Abelian magnon potential $\B_\mu$. Let $(\n_1,\n_2,\n_3)$
be an arbitrary orthonormal frame in the $SU(2)$ 
space, and choose an arbitrary direction $\n$ as 
the Abelian direction and let $\n=\n_3$. With this we can decompose $\B_\mu$ to the restricted potential $\hB_\mu$ and the valence potential $\W_\mu$ \cite{prd80,prl81}
\begin{gather}
\B_\mu= \hB_\mu +\W_\mu,~~~\n \cdot \W_\mu=0,  \nn\\
\B_\mu= B_\mu \n -\frac1g \n \times \pd_\mu \n,
~~~B_\mu =\n \cdot \B_\mu,
\label{cdec}
\end{gather}
where $\hB_\mu$ is the Abelian projection of 
$\B_\mu$ fixed by the isometry condition
\begin{gather}
D_\mu \n =0.
\label{ap}
\end{gather}
Notice that $\hB_\mu$ is the potential which 
makes the Abelian direction a covariant constant, 
the potential which parallelizes $\n$.

The Abelian decomposition has the following 
features \cite{prd80,prl81}. First, the restricted potential has a dual structure. It is made of two parts, the non-topological (Abelian) Maxwellian 
part $B_\mu \n$ and topological Diracian part
$(1/g) \n \times \pd_\mu \n$. Second, $\hB_\mu$ retains the full non-Abelian gauge degrees of freedom, while $\W_\mu$ transforms gauge covariantly. Finally, the decomposition is gauge independent. Once the Abelian direction is chosen, it follows automatically, independent of the gauge. And we can choose any direction as the Abelian direction.   

From the Abelian decomposition (\ref{cdec}) 
we have
\begin{gather}
\G_\mn=\hG_\mn + \hD _\mu \W_\nu 
- \hD_\nu \W_\mu + g' \W_\mu \times \W_\nu,   \nn\\
\hD_\mu=\pd_\mu+g' \hB_\mu \times,   \nn\\
\hG_\mn= \pd_\mu \hB_\nu-\pd_\nu \hB_\mu
+ g \hB_\mu \times \hB_\nu =G_\mn' \n, \nn \\
G'_\mn=G_\mn + H_\mn
= \pd_\mu B_\nu'-\pd_\nu B_\mu',  \nn\\
G_\mn =\pd_\mu B_\nu-\pd_\nu B_\mu, \nn\\
H_\mn =-\frac1{g'} \n \cdot (\pd_\mu \n 
\times\pd_\nu \n) =\pd_\mu C_\nu-\pd_\nu C_\mu, \nn\\
B_\mu' = B_\mu+ C_\mu,
~~~C_\mu =-\frac1{g'} \n_1\cdot \pd_\mu \n_2.
\end{gather}
Notice that the restricted field strength $\hG_\mn$ inherits the dual structure of $\hB_\mu$, so that 
it can also be described by two Abelian potentials, the Maxwellian $B_\mu$ and the Diracian $C_\mu$. 
But here the potential $C_\mu$ for $H_\mn$ is determined uniquely up to the $U(1)$ gauge freedom which leaves $\n$ invariant. This is the Abelian decomposition of the SU(2) gauge field known as 
the Cho decomposition, Cho-Duan-Ge (CDG) decomposition, or Cho-Faddeev-Niemi (CFN) 
decomposition \cite{fadd,shab,zucc,kondo}. 

Expressing the charged spinon doublet $\phi$ by 
the scalar Higgs field $\rho$ and the SU(2) unit doublet $\xi$ by
\begin{gather}
\phi = \frac{1}{\sqrt{2}} \rho~\xi,
~~~(\xi^\dag \xi = 1), 
\label{xi}
\end{gather}
and using the Abelian decomposition (\ref{cdec})
we have
\begin{gather}
\cD_\mu \xi= \Big[\pd_\mu-i\frac{g}{2} A_\mu 
-i\frac{g'}{2} (B_\mu' \n +\W_\mu) \cdot \vsig \Big]~\xi,  \nn\\
|\cD_\mu \xi|^2 =\frac{1}{8} (-gA_\mu+g'B_\mu')^2 
+\frac{g'^2}{4} \W_\mu^2.
\end{gather}
From this we can remove the SU(2) unit doublet $\xi$ completely from the Lagrangian and ``abelianize" it gauge independently \cite{prd80,prl81}
\begin{gather}
\cL = -\frac12 (\pd_\mu \rho)^2
-\frac{\lam}{8}\big(\rho^2-\rho_0^2 \big)^2 \nn\\
-\frac14 F_\mn^2 -\frac14 {G_\mn'}^2
-\frac12 \big|D_\mu' W_\nu 
-D_\nu' W_\mu \big|^2  \nn\\	
-\frac{\rho^2}{8} \big((-gA_\mu+g'B_\mu')^2 
+2 g'^2 W_\mu^*W_\mu \big)  \nn\\
+i g' G_\mn' W_\mu^* W_\nu 
+ \frac{g'^2}{4}(W_\mu^* W_\nu 
-W_\nu^* W_\mu)^2,  \nn\\
D_\mu'=\pd_\mu +ig' B_\mu',
~~~W_\mu =\frac{1}{\sqrt 2} (W^1_\mu +i W^2_\mu),
\label{lag1}
\end{gather}
where $\rho_0=\sqrt{2\mu^2/\lambda}$. This tells that the Lagrangian (\ref{lag0}) is made of two Abelian gauge potentials, the electromagnetic $A_\mu$ and the magnonic Abelian $B_\mu'$, 
a complex magnon $W_\mu$, and the Higgs scalar 
field $\rho$. 

{\bf Photon-magnon mixing}---To understand what happened to $\xi$, notice that the two Abelian 
gauge fields in the Lagrangian are not mass eigenstates. To express them in terms of mass eigenstates, we introduce the following photon-magnon mixing by
\begin{gather}
\left(\begin{array}{cc} \bA_\mu \\
Z_\mu  \end{array} \right)
=\frac{1}{\sqrt{g^2 +g'^2}} \left(\begin{array}{cc} 
g' & g \\ -g & g' \end{array} \right)
\left(\begin{array}{cc} A_\mu \\ B'_\mu
\end{array} \right)  \nn\\
= \left(\begin{array}{cc}
\cos \om & \sin \om \\
-\sin \om & \cos \om \end{array} \right)
\left(\begin{array}{cc} A_\mu \\ B_\mu'
\end{array} \right),
\label{mix}
\end{gather}
where $\om$ is the mixing angle. With this we can 
express the Lagrangian (\ref{lag1}) by
\begin{gather}
\cL = -\frac12 (\pd_\mu \rho)^2
-\frac{\lam}{8}\big(\rho^2-\rho_0^2 \big)^2
-\frac14 {\bF_\mn}^2 -\frac14 Z_\mn^2 \nn\\
-\frac12 \big|(\bD_\mu +i\be\frac{g'}{g}Z_\mu)W_\nu 
-(\bD_\nu +i \be\frac{g'}{g} Z_\nu)W_\mu \big|^2  \nn\\
-\frac{\rho^2}{4} \big(g'^2 W_\mu^*W_\mu
+\frac{g^2+g'^2}{2} Z_\mu^2 \big) +i \be (\bF_\mn 
+\frac{g'}{g}  Z_\mn) W_\mu^* W_\nu   \nn\\
+ \frac{g'^2}{4}(W_\mu^* W_\nu -W_\nu^* W_\mu)^2,
\label{lag2}
\end{gather}
where 
\begin{gather}
\bF_\mn=\pd_\mu \bA_\nu-\pd_\nu \bA_\mu, 
~~~Z_\mn = \pd_\mu Z_\nu-\pd_\nu Z_\mu,  \nn\\
\bD_\mu=\pd_\mu+i \be \bA_\mu,   \nn\\
\be=\frac{gg'}{\sqrt{g^2+g'^2}}=g' \sin\om =g \cos\om.
\label{e}
\end{gather}	
This is the physical expression of the non-Abelian magnon spintronics Lagrangian (\ref{lag0}), which tells that the three degrees of $\xi$ are absorbed to $Z_\mu$ and $W_\mu$ to make them massive, so that the theory is made of Higgs scalar $\rho$, massless and massive Abelian gauge bosons $\bA_\mu$ and $Z_\mu$, and massive complex $W_\mu$ magnon whose masses are given by 
\begin{gather}
M_H= {\sqrt \lam} \rho_0,  \nn\\ 
M_W=\frac{g'}{2} \rho_0,
~~~M_Z=\frac{\sqrt {g^2+g'^2}}{2} \rho_0.		
\end{gather} 
So it has three mass scales. 

From the physical point of view this Lagrangian
looks totally different from the original
Lagrangian (\ref{lag0}). In particular, 
the $SU(2)\times U(1)$ gauge symmetry seems to have disappeared completely here. But we emphasize 
that the Lagrangian (\ref{lag2}) is mathematically identical to (\ref{lag0}), so that it retains 
the full non-Abelian gauge symmetry of the original Lagrangian. It is hidden, but not disappeared. 

The mixing (\ref{mix}) has deep implications. 
First, notice that the two Abelian gauge bosons $\bA_\mu$ and $Z_\mu$ in (\ref{lag2}) could naturally be interpreted to represent the real photon and the neutral magnon. This immediately tells that the massive magnon $W_\mu$ carries 
the electric charge as well as the spin charge, 
so that (just like the electron doublet) it is doubly charged. This must be clear from (\ref{lag2}), which shows that $W_\mu$ couples to both $\bA_\mu$ and $Z_\mu$. Obviously, $W_\mu$ acquires the electric charge from $\xi$ absorbing the charge carried by the unit doublet. 

Now, we may ask which of $\bA_\mu$ and $Z_\mu$ describes the real photon. The answer depends on whether we have a long range magnetic order 
in the spintronics or not. Since the long range magnetic order is a crucial concept in spintronics which is essential for many spintronic devices, 
we may assume that the long range magnetic order does exist in spintronics. If so, we must identify the massless $\bA_\mu$ as the massless magnon 
which is responsible for the long range magnetic oder, and identify $Z_\mu$ as the massive photon. 
In this case the coupling constant of $Z_\mu$ 
which couples to $W_\mu$ in (\ref{lag2}) should become the electric charge,
\begin{gather}
\be \frac{g'}{g} =e.
\end{gather}  
From this we have (with $g/2=e$) 
\begin{gather}
g'= \frac{\sqrt{1+\sqrt{17}}}{\sqrt 2} e,
~~~\tan \om= \frac{2\sqrt 2}{\sqrt{1+\sqrt{17}}},
\end{gather} 
so that 
\begin{gather}
M_W =\frac{\sqrt{1+\sqrt{17}}}{2\sqrt 2} e \rho_0, \nn\\
M_Z =\frac{\sqrt{9+\sqrt{17}}}{2\sqrt 2} e \rho_0  \simeq 1.6 M_W.
\label{mass2}
\end{gather}
This tells that the penetration length of 
the off-diagonal magnon field is 1.6 times bigger than that of the photon. More importantly, 
the massless Abelian magnon generates a long 
range magnetic interaction in the non-Abelian 
spintronics.

With this interpretation we can express 
the Lagrangian (\ref{lag2}) in the final form,
\begin{gather}
\cL = -\frac12 (\pd_\mu \rho)^2
-\frac{\lam}{8}\big(\rho^2-\rho_0^2 \big)^2
-\frac14 {\bF_\mn}^2  \nn\\ 
-\frac14 Z_\mn^2 
-\rho^2 \frac{M_Z^2}{2\rho_0^2}  Z_\mu^2 \nn\\
-\frac12 \big|(\bD_\mu +ie Z_\mu)W_\nu 
-(\bD_\nu +ie Z_\nu)W_\mu \big|^2   \nn\\
+i (\be~\bF_\mn + e~Z_\mn) W_\mu^* W_\nu   \nn\\ 
- \frac{M_W^2}{\rho_0^2} \Big[\rho^2 W_\mu^*W_\mu -(W_\mu^* W_\nu -W_\nu^* W_\mu)^2 \Big],  \nn\\
\be = \frac{2\sqrt 2}{\sqrt{1+\sqrt{17}}}~e
\simeq 1.25~e.
\label{lag3}
\end{gather}
The contrast between this and (\ref{lag0}) is unmistakable. This tells that the magnonic spin coupling $\be$ 
is completely fixed by $e$, so that the theory has 
only one coupling constant. Notice that the spin coupling is stronger than the electromagnetic 
coupling.

At this point we need to discuss the Higgs mechanism 
to point out a common misunderstanding on this mass generation mechanism. The popular explanation of 
the Higgs mechanism is that a spontaneous symmetry breaking by the vacuum $\langle \phi \rangle$ of 
the Higgs doublet generates the mass. But obviously 
here the mass generation of $Z_\mu$ and $W_\mu$ in (\ref{lag2}) comes from the non-vanishing vacuum value of the scalar field $\rho$, not by 
$\langle \phi \rangle=\rho_0 \langle \xi \rangle$. And as a scalar $\rho_0$ can NOT break any symmetry, spontaneous or not. Perhaps more importantly, 
here $\langle \xi \rangle$ could still fluctuate 
at the vacuum with the Higgs mechanism, so that 
we have the mass generation without any symmetry breaking. This tells that the popular Higgs mechanism by spontaneous symmetry breaking is only half of the full story \cite{pla23,ap24,arx25}.

{\bf Conserved charge and spin currents and their mixing}---In spintronics the spin current is 
an important ingredient which plays a central role, so that one may ask how the above theory describes the spin current. To answer this notice that we 
have the following equations of motion from 
the Lagrangian (\ref{lag3}), 
\begin{gather}
\pd^2 \rho-\Big(2\frac{M_W^2}{\rho_0^2} W_\mu^*W_\mu 
+ \frac{M_Z^2}{\rho_0^2}~Z_\mu^2 \Big)~\rho
=\frac{\lambda}{2}\big (\rho^2 
-\rho_0^2 \big)~\rho,   \nn\\	
\Big(\bD_\mu+ ie Z_\mu \Big) 
\Big[(\bD_\mu +ie Z_\mu) W_\nu 
-(\bD_\nu +ie Z_\nu) W_\mu \Big]  \nn\\
=i e W_\mu \big(\bF_\mn + Z_\mn \big)
+M_W^2 \frac{\rho^2}{\rho_0^2} W_\nu   \nn\\
+ 4\frac{M_W^2}{\rho_0^2} W_\mu(W_\mu^* W_\nu -W_\nu^* W_\mu), \nn\\	
\pd_\mu \Big[\bF_\mn -i\be(W_\mu^* W_\nu
-W_\nu^* W_\mu) \Big]  \nn\\
=i \be \Big[W_\mu^* (\bD_\mu W_\nu -\bD_\nu W_\mu) -(\bD_\mu W_\nu -\bD_\nu W_\mu)^* W_\mu \Big]  \nn\\
+ e \be \Big[2 W_\mu^*W_\mu Z_\nu
-Z_\mu(W_\mu^*W_\nu +W_\nu^*W_\mu) \Big],  \nn\\
\pd_\mu \Big[Z_\mn-ie (W_\mu^* W_\nu
- W_\nu^* W_\mu ) \Big] 
-M_Z^2 \frac{\rho^2}{\rho_0^2} Z_\nu  \nn\\
=ie \Big[W_\mu^*(\bD_\mu W_\nu -\bD_\nu W_\mu) 
-W_\mu (\bD_\mu W_\nu -\bD_\nu W_\mu)^* \Big]  \nn\\
+e^2 \Big[ 2W_\mu^*W_\mu Z_\nu
-Z_\mu(W_\mu^*W_\nu + W_\nu^*W_\mu)\Big].
\label{eom}
\end{gather}
Clearly the last two equations can be put 
in the form
\begin{gather}
\pd_\mu \bF_\mn =J_\nu^{(s)},  
~~~~\pd_\mu Z_\mn =J_\nu^{(e)},  
\label{cj}
\end{gather}
where $J_\mu^{(s)}$ and $J_\mu^{(e)}$ are the spin and charge currents which correspond to the Abelian potentials $\bA_\mu$ and $Z_\mu$. This proves that the theory has two conserved currents.

The existence of the two conserved currents originates from the existence of two Abelian gauge symmetries of the original Lagrangian (\ref{lag0}), the electromagnetic $U(1)$ and the $U(1)$ subgroup of $SU(2)$ which leaves the Abelian direction $\n$ invarint. This must be clear in (\ref{lag1}), which has two Abelian potentials $A_\mu$ and $B'_\mu$. 
So it has two conserved currents, 
the electromagnetic current $j_\mu^{(e)}$ of $A_\mu$ 
and the magnonic spin current $j_\mu^{(s)}$ of $B_\mu'$. These two currents mix together because of the mixing (\ref{mix}), and we have the two physical conserved currents shown in (\ref{cj}) after 
the mixing. 

This tells two things. First, both $J_\mu^{(s)}$ 
and $J_\mu^{(e)}$ contain the electromagnetic 
current of $j_\mu^{(e)}$ of $A_\mu$ and the spin 
current of $j_\mu^{(s)}$ of $B_\mu'$. Second, both 
(in particular, $J_\mu^{(s)}$) contain the electron 
spin current and magnon spin current. This could 
explain the conversion of the spin and charge 
currents as well as the interconversion between 
the electron spin current and magnonic spin current 
in spintronics \cite{mend,cap}.    

The importance of the conserved charge and spin 
currents in spintronics has been emphasized in 
the literature before \cite{chumak,jmmm,hb}. As far as we understand, however, there have been no 
theory of spintronics which could show where 
the two currents come from, what are they made of, and how we can express them so far. For the first time, we have a theory of spintronics which could answer these question completely. With this we 
could study the role of these currents, for 
instance the possible interconversion of the twe currents, in spintronics in more detail.   

{\bf Discussions}---In this Letter we have constructed a non-Abelian $SU(2)\times U(1)$ magnonic gauge theory of electron spintronics, treating the electron as a charged spinon. 
The theory has the following charactristic features. First, it has the photon-magnon mixing which produces a massless Abelian magnon which explains 
the existence of a long range magnetic order in spintronics. Second, the same photon-magnon mixing makes the massive non-Abelian magnon doubly 
charged, so that the non-Abelian magnon which induces the spin-flip interaction carries the electric (as well as the spin) charge. Third, it 
has the Higgs mechanism which makes the photon 
and the non-Ableian magnon massive, without any spontaneous symmetry breaking. Fourth, it has 
the non-Abelian Meissner effect generated by 
the charged magnon as well as the Abelian Meissner effect generated by the massive photon. Fifth, it has two conserved currents, the electromagnetic 
and spin currents. So the theory has all necessary ingredients that we need to describe the electron spintronics.   

Notice that when we switch off the massive photon $Z_\mu$, the Lagrangian (\ref{lag2}) reduces to 
\begin{gather}
\cL = -\frac12 (\pd_\mu \rho)^2
-\frac{\lam}{8}\big(\rho^2-\rho_0^2 \big)^2
-\frac14 {\bF_\mn}^2   \nn\\
-\frac12 \big|\bD_\mu W_\nu -\bD_\nu W_\mu \big|^2 
+i \be~\bF_\mn W_\mu^* W_\nu  \nn\\
- \frac{M_W^2}{\rho_0^2} \Big[\rho^2 W_\mu^*W_\mu -(W_\mu^* W_\nu -W_\nu^* W_\mu)^2 \Big],
\label{lglag}
\end{gather}
which describes the non-Abelian magnon gauge interaction in the frustrated magnetic materials proposed recently by Zarzuela and Kim \cite{zarkim}. Moreover, when we swtich off the three magnons $\bA_\mu$ and $W_\mu$, it reduces to the well 
known Abelian Landau-Ginzburg Lagrangian of 
the ordinary superconductors \cite{lg}. This strongly implies that the physics of the spintronics could be closely related to the physics of 
the strongly correleted magnetism and/or the spin doublet superconductivity. 

How closely are they related? Very closely \cite{arx25}. There are many indications for this, which has made the superconducting spintronics 
an important part of spintronics \cite{lind,esch}. But most of them are concerened with spin-triplet superconductors. Our work in this letter strongly implies that the physics of the electron spintronics is not just closely related to that of the two-gap ferromagnetic superconductivity, but is almost identical. 

This is because the Lagrangian (\ref{lag0}) is formally identical to the Lagrangian that we have proposed to describe the two-gap ferromagnetic superconductivity recently \cite{pla23,ap24,arx25}. The only difference is that in the two-gap ferromagnetic superconductivity the $SU(2)$ doublet $\phi$ describes the spin doublet Cooper pair which 
carries the charge $2e$. But here $\phi$ describes the charged spinon which carries the charge $e$. 
So they describe different physical phenomena. Nevertheless, mathematically they are identical.
This assures that the two theories have exactly 
the same underlying physics.

An important charactristic feature of these 
theories is the existence of non-Abelian 
topological objects, in particular the quantized magnonic vortex and monopole, as well as 
the Abelian Abrikosov type magnetic vortex. 
Obviously the existence of these topological 
objects comes from the non-Abelian $\pi_1(S^1)$ topology and $\pi_2(S^2)$ topology of 
the Lagrangian (\ref{lag0}). Since this 
Lagrangian also describes the two-gap ferromagnetic superconductivity, it must be clear the above 
theory of spintronics has all topological 
objects which we have in two-gap ferromagnetic superconductors \cite{pla23,ap24,arx25}. In particular, the spontronics theory has the quantized magnonic vortex which has the spin flux $2\pi/\be$ with a non-trivial Higgs and $W_\mu$ dressing 
and the Cho-Maison type magnonic monopole which carries the spin magnetic charge $4\pi/\be$, 
which can be viewed as a hybrid between Dirac and 'tHooft-Polyakov monopole \cite{thooft,plb97}.  

Most importantly, it should be emphasized that 
the Lagrangian (\ref{lag0}) is precisely 
the Weinberg-Salam Lagrangian which is well 
known to describe the standard model in high 
energy physics \cite{pla23,ap24,arx25}. With 
the experimental confirmation of the Higgs 
particle at LHC, the standard model has become 
a most successful theory which unifies the weak 
and electromagnetic interactions in high energy physics. And here we propose the same theory as 
the theory of the electron spintronics. This 
is surprising.

Of course, the Lagrangian originally proposed by 
Weinberg has totally different meaning here 
in spintronics \cite{wein}. For instance, 
in the standard model the massless $\bA_\mu$ 
describes the real massless photon, but in 
the above spintronics it describes the massless 
magnon. So the Cho-Maison monopole in the standard model is a real electromagnetic monopole which carries the magnetic charge $4\pi/e$, but here in spintronics the Cho-Maison monopole is a magnonic (not electromagnetic) monopole as we have pointed out. Perhaps more importantly, the energy scale 
in two theories is totally different. In fact, 
the Higgs vacuum value in the standard model is 
of the order of 100 GeV, but here in spintronics 
is supposed to be of the order of meV, different 
by the factor $10^{14}$. So they describe totally different physics in totally different surroundings. Nevertheless, they are mathematically identical. 
It is really remarkable that the same theory could describe totally different physics in condensed matters.  

This brings us to an important issue. Since 
the standard model predicts the Cho-Maison 
monopole, the experimental confirmation of 
the Cho-Maison monopole has been regarded as 
the final and topological test of the standard 
model. For this reason MoEDAL and ATLAS at 
LHC have been actively searching for 
the monopole \cite{med1,med2,med3,atlas}.
But the above discussion strongly implies that 
we could also have the same monopole in condensed 
matters in spintronic materials and/or spin 
doublet superconductors. This makes the search 
for the Cho-Maison monopole in condensed matters 
an important task. Although the mass of 
the Cho-Maison monopole in the standard model 
is expected to be around 11 TeV, the mass in 
condensed matters would be much less, around 
100 meV. It would be really interesting if we 
can confirm the existence of such monopole 
in condensed matters in table top experiment, competing with the biggest accelerator in 
the world.

The spintronics is a very important branch of 
condensed matter physics which has potentially 
huge applications in data processing and computing devices. In this Lettter we have proposed a new $SU(2) \times U(1)$ gauge theory of the electron spintronics where the spin-spin interaction is explained by the exchange of the messenger bosons. This and similar recent works on the non-Abelian gauge interactions in the frustrated magnetic materials and multi-gap superconductors strongly indicates that the non-Abelian gauge interaction mediated by the messenger bosons could become 
a main stream in condensed matter physics in 
the future \cite{pla23,ap24,arx25,zarkim}. 

We hope that our work could play important roles 
in our understanding of the magnon spintronics, 
and become a starting point for the realistic 
theory of spintronics. The details of the theory, 
in particular the conversion of the charge and 
spin currents and the generalization of the theory to the spin-triplet spintronics, will be discussed 
in a separate paper \cite{cho}.

{\bf ACKNOWLEDGEMENT}

~~~The work is supported by the National Research Foundation of Korea funded by the Ministry of Science and Technology (Grant 2025-R1A2C1006999) and by Center for Quantum Spacetime, Sogang University, Korea.

\end{document}